# Acousto-drag photovoltaic effect by piezoelectric integration of two-dimensional semiconductors

Jiaming Gu[1,#], Yicheng Mou[1,#], Jianwen Ma[1,#], Haonan Chen[1,#], Chuanxin Zhang[2], Yuxiang Wang[1], Jiayu Wang[1], Hangwen Guo[1,3,4], Wu Shi[1,3], Xiang Yuan[5,6], Xue Jiang[2], Dean Ta[2], Jian Shen[1,3,4,7], Cheng Zhang[1,3*]

[1] *State Key Laboratory of Surface Physics and Institute for Nanoelectronic Devices and Quantum Computing, Fudan University, Shanghai 200433, China*

[2] *Center for Biomedical Engineering, School of Information Science and Technology, Fudan University, Shanghai 200433, China*

[3] *Zhangjiang Fudan International Innovation Center, Fudan University, Shanghai 201210, China*

[4] *Shanghai Qi Zhi Institute, Shanghai 200232, China*

[5] *State Key Laboratory of Precision Spectroscopy, East China Normal University, Shanghai 200241, China*

[6] *School of Physics and Electronic Science, East China Normal University, Shanghai 200241, China*

[7] *Collaborative Innovation Center of Advanced Microstructures, Nanjing 210093, China*

[#] *J.G., Y.M., J.M. and H.C. contributed equally to this work*

[*] Correspondence and requests for materials should be addressed to C. Z. (E-mail: zhangcheng@fudan.edu.cn)

**Abstract**

Light-to-electricity conversion is crucial for energy harvesting and photodetection, requesting efficient electron-hole pair separation to prevent recombination. Traditional junction-based mechanisms using built-in electric fields fail in non-barrier regions. Homogeneous material harvesting under photovoltaic effect is appealing but only realized in non-centrosymmetric systems via bulk photovoltaic effect. Here we report the realization of photovoltaic effect by employing surface acoustic waves (SAW) to generate zero-bias photocurrent in a conventional layered semiconductor $MoSe_2$. SAW induces periodic modulation to electronic bands and drags the photoexcited pairs toward the travelling direction. The photocurrent is extracted by a local barrier. The separation of generation and extraction processes suppresses recombination and yields large nonlocal photoresponse. We distinguish acousto-electric drag and electron-hole pair separation effect by fabricating devices of different configurations. The acousto-drag photovoltaic effect, enabled by piezoelectric integration, offers an efficient light-to-electricity conversion method, independent of semiconductor crystal symmetry.

The conversion of light into electricity plays a central role in modern technologies, ranging from solar energy harvesting, video imaging, to optical communications[1–4]. An important step towards high photoresponse is the efficient separation of photoexcited electron-hole pairs before recombination. It is usually achieved by build-in electric fields in various junction-type architectures such as *p-n* diodes and Schottky diodes. In these devices, the photoexcited pairs are separated and transported to opposite directions simultaneously by local potential barriers, while the photocurrent vanishes rapidly outside the barrier region. Generating photovoltaic effect in a homogeneous material upon uniform illumination could help to develop new optoelectronic devices beyond the widely used junction-type design, and has been intensively investigated over these years[5–13]. One of the few ways to do so is through the bulk photovoltaic effect (BPVE)[14,15], which results from the asymmetric excitation or scattering of photoexcited electron–hole pairs in materials without inversion symmetry[16–18]. Notable BPVE systems include ferroelectric insulators and semiconductors such as perovskite oxides/halides[15], GeTe[19], and $CuInP_2S_6$[20], topological Weyl semimetal TaAs[21], magnetic insulator $CrI_3$[22], polar $WS_2$ nanotubes[6], and so on. However, the restriction on crystal symmetry strongly limits the application of BPVE in the broad range of materials with high photoresponse.

Here we realize a new design of photovoltaic effect, termed as acousto-drag photovoltaic effect (ADPVE), which yields large nonlocal photoresponse in the bulk of a homogeneous material. In contrast to the BPVE-type photocurrent generation crucially relying on crystal symmetry breaking, ADPVE proposed here employs a technique called surface acoustic wave (SAW) to locally separate the photoexcited electron–hole pairs by periodic band modulation, and drag them toward current-collecting electrodes. The Rayleigh-type SAW is used, which travels along the surface of an elastic solid medium with a shallow penetration depth and elliptical polarization[23–25]. Experimentally, SAW can be excited by high-frequency electric signals through interdigital transducers (IDTs) fabricated on a piezoelectric substrate like $LiNbO_3$. The periodicity of IDTs is designed to match the preconceived frequency and the acoustic velocity of the substrate[24,25]. Despite the demonstration of SAW-excited carrier transport decades ago, the difficulty of introducing piezoelectricity in semiconductors limits its application to very few material systems[26–28]. While widely used as delay lines, filters and oscillators in electronic circuit applications, SAW can be used to probe and control elementary excitations such as electrons, magnons, and excitons in condensed matter physics in recent years[29–41]. In particular, the feasible integration of van der Waals materials on piezoelectric substrates opens a new frontier of electronic and optoelectronic research with acoustic tunability.

In Fig. 1a-c, we present the comparison of the photocurrent generation processes in conventional semiconductors without (a) and with (b) band bending induced by electric fields, and with the application of SAW (c). In the absence of band bending (Fig. 1a), the photoexcited electron-hole pairs rapidly recombine without producing photocurrent. When a band bending is induced by external bias voltage or internal charge polarization, the photoexcited electron-hole pairs will be split due to the opposite drifting directions of electrons and holes (Fig. 1b). A net photocurrent is generated in the band-bending systems like *p-n* junctions or ferroelectric materials. Except band bending, other mechanisms like shift current have also been used to generate zero-bias photocurrent. Distinct from these scenarios, ADPVE proposed here uses SAW to generate periodically modulated electronic bands in an adjacent van der Waals semiconductor, like $MoSe_2$ as we used in this work (Fig. 1c-d). The conduction and valence bands are at the same spatial phase, known as type-II band modulation[42], as opposite to the type-I band modulation, in which the maxima of conduction and valence bands are shifted by half a wavelength of SAW[43]. When illuminated, the photoexcited electrons and holes will be spatially separated to the minima of conduction band and the maxima of valence band. These electrons and holes are collectively dragged along the travelling direction of SAW while the spatial separation effectively prevents them from recombination. At the end of the semiconductor channel, a metal contact is used to extract the photocurrent through local Schottky barrier. ADPVE discussed here originates from the model proposed by Yakovenko et al. in 2012[44]. Instead of using limited piezoelectric semiconductors which typically have lower electromechanical transduction efficiency, we integrate layered van der Waals semiconductors on the top of piezoelectric insulators to achieve both high photosensitivity and high electromechanical coupling coefficient simultaneously. Note that unlike other photocurrent mechanisms, here the pair generation and photocurrent extraction are two independent processes, which allows for large nonlocal directional photocurrent generated in the bulk.

Experimentally, we fabricated our devices on commercially-available 128° Y-cut $LiNbO_3$ substrates consisting of pre-patterned pairs of IDTs by photolithography and metal sputtering. Fig. 1d presents the IDT we designed with a periodicity of 13.2 μm for 300 MHz center frequency and metallization ratio of 0.33. Fig. 1e shows the simulated electric potential profile induced by SAW, showing clear periodic oscillations and highly directional propagation. The in-plane and out-of-plane piezoelectric field components ($E_x$ and $E_z$) are calculated in Fig. 1e. Here the presence of SAW mainly induces type-II band modulation from the piezoelectric field[45]. Exfoliated semiconducting 2H-$MoSe_2$ nanoflakes were transferred onto the central region of an IDT pair. Two Au/Cr (75/5 nm) electrodes were deposited at both ends of the channel, forming two opposite Schottky junctions. The sample size (10-25 μm) is much smaller than the aperture of IDTs (260 μm), ensuring a uniform acoustic intensity across the channel. Vector Network Analyzer (VNA) was applied to examine the electromechanical transduction efficiency and determine the center frequency. Fig. 1f is the measured reflection and transmission coefficients ($S_{11}$ and $S_{21}$) between a pair of IDTs as a function of frequency. $S_{11}$ presents a sharp valley (-15 dB) near 285 MHz where a corresponding peak in $S_{21}$ (-13 dB) appears.

We firstly studied the photovoltage response of a $MoSe_2$/$LiNbO_3$ device (Device 1) in a transverse configuration, where SAW travels perpendicularly to the direction between two contact electrodes. In this configuration, the detected photocurrent is independent of ADPVE discussed above for the orthogonality to each other. A 633 nm laser was focused on $MoSe_2$ flakes with a spot diameter below 1.5 μm and modulated by a programmed scanning mirror. The open-circuit photovoltage ($V_{OC}$) is measured by lock-in amplifiers when laser is chopped at 200-300 Hz. Fig. 2a is the photovoltage mapping of $MoSe_2$/$LiNbO_3$ device (Device 1) collected between the left and right electrodes. The optical image of the measured device is shown in Fig. 2b, in which the black dotted box is the corresponding scanning area. In the $V_{OC}$ mapping (Fig. 2a), the device generates opposite photovoltages near two contact electrodes. As SAW is on, the photovoltages on both sides get enhanced with the increase of acoustic intensity, while the photovoltage in the middle remains nearly zero. The photocurrent mapping shows similar patterns when reversing the SAW travelling direction. The underlying mechanism is illustrated in Fig. 2c. The band modulation induced by SAW results in the spatial separation of photoexcited electron-hole pairs, helping to suppress recombination. Hence it leads to enhancement of photovoltage response. The detailed line-profile analysis in Fig. 2d shows that the position dependence of $V_{OC}$ remains similar for upward- and downward-travelling SAW comparing to that of SAW-OFF state. Fig. 2e summarizes the acoustic intensity $I_{SAW}$ dependence of $V_{OC}$ and short-circuit photocurrent $I_{SC}$, which is symmetric with respect to the reversal of SAW direction. The acoustic intensity is extracted using input RF power and $S_{21}$, considering the acoustic-electric symmetric conversion of an IDT pair. The arrows of acoustic intensity refer to the opposite direction of SAW. Considering an intrinsic semiconductor, we simulated the energy band and carrier concentration distribution of 2H-MoSe$_2$. Consistent with our expectations, the energy band is periodically modulated as type-II band modulation (see Fig. S7), where electrons tend to shift to the minima of conduction band while holes move to the maxima of valence band[46]. These results further confirm the electron-hole separation effect. The electron-hole separation effect is independent of the SAW direction and applies to the longitudinal configuration as well.

We then investigated the spatial distribution of photovoltage response in a longitudinal configuration, where SAW travels parallelly to the direction between two contact electrodes. Fig. 3a shows the spatial mapping of $V_{OC}$ (Device 2). The optical image is presented in Fig. 3b, in which SAW is parallel to the measured photocurrent direction. Without SAW, the region with high $V_{OC}$ response mainly locates near contact electrodes, where Schottky barrier is present, while negligible $V_{OC}$ signal in the center part (Fig. 3a). The signs of $V_{OC}$ near two electrodes are opposite due to the reversed band bending directions. When rightward SAW is applied, a strong negative $V_{OC}$ signal appears in the middle with a much larger absolute value than that without SAW. Upon the increase of the SAW power, $|V_{OC}|$ shows dramatic enhancement with high-response region shifting towards the center of the sample. In contrast, a positive $V_{OC}$ signal appears in the center region when reversing the direction of SAW (Fig. 3a). Fig. 3a suggests that the emerging nonlocal photovoltage in the center results from the acousto-electric drag effect of travelling SAW, which carries the photoexcited electron-hole pairs to the local Schottky barrier induced by contact electrodes (Fig. 3c), resulting in the enhancement of $|V_{OC}|$. Fig. 3e summarizes the acoustic intensity $I_{SAW}$ dependence of $V_{OC}$ and $I_{SC}$, which scale linearly with $I_{SAW}$. The linearity of photoresponse with $I_{SAW}$ shares similar origins with the standard acousto-electric effect.[27] The slight difference in photocurrent value between leftward- and rightward-travelling SAW results from the variations of electromechanical transduction efficiency among different IDTs and Schottky barriers of different electrodes. In Fig. 3d, the line profile of $V_{OC}$ (obtained from the red dashed line in Fig. 3b) only shows clear photoresponse near the electrode edges when SAW is absent (the middle panel in Fig. 3d). However, when applying SAW, such edge response profile gradually gets overwhelmed by the rise of a large peak from the center region (top and bottom panels in Fig. 3d). The dramatical difference in the line profile is clear evidence for different photocurrent generation mechanisms for SAW-ON and OFF states. The presence of SAW effectively collects the photoexcited electron-hole pairs in the center part, which otherwise would recombine very quickly, and converts them into photocurrent. The generated photocurrent is most prominent in the middle and slightly decays near the left and right edges since part of the laser spot may lie outside the channel. And the photovoltage enhancement from electron-hole separation (Fig. 2e) is much weaker than that of the acousto-electric drag effect (Fig. 3e). To exclude the influence of the acousto-electric voltage caused by the net charge carriers in the sample, we employ back gate to modulate the carrier concentration and carrier type in the MoSe$_2$ channel through LiNbO$_3$ substrate (Fig. S11). The conventional acousto-electric effect arises from the directional motion of net charge carriers along with SAW. In contrast, ADPVE involves the separation and propagation of photoexcited electron-hole pairs under SAW, as well as the generation of photocurrent through Schottky junctions.[27,29] ADPVE persists at the insulating region with negative gate voltage, where the acousto-electric voltage becomes negligible. It suggests that ADPVE is independent from possible acousto-electric effect.

To characterize the device photodetector performance, we calculate typical photoresponse parameters including responsivity ($R_{PC}=I_{SC}/P_{laser}$ and $R_{PV}=V_{OC}/P_{laser}$, $I_{SC}$ is the short-circuit photocurrent, $V_{OC}$ is the open-circuit photovoltage, $P_{laser}$ is the input laser power), external quantum efficiency (EQE, $\eta=n_e/n_{photon}=(hcI)/(e\lambda P_{laser})$, $h$ is the Planck constant, $c$ is the light speed, $I$ is the photocurrent, $e$ is the elementary electron charge, $\lambda$ is the laser wavelength, $P_{laser}$ is the input laser power), and specific detectivity ($D^*=(A\Delta f)^{1/2}/NEP$, $A$ is the area of the photosensitive region of the detector, $\Delta f$ is the bandwidth, and NEP is the noise equivalent power). As shown in Fig. 4a (Fig. S14), we measured the SAW-enhanced $V_{OC}$ and $I_{SC}$ at different laser powers. $V_{OC}$ and $I_{SC}$ increase rapidly with the increase of acoustic intensity. The enhancement of $V_{OC}$ and $I_{SC}$ are higher under weak light intensity while gradually decreasing due to enhanced recombination rate and screening of the piezoelectric field at higher laser power. The variations between responsivity and input laser power are provided in Fig. 4b. We find that the responsivity can be enhanced with SAW by a ratio of 828. For the same acoustic intensity, both PC and PV responsivity is better at weak light intensity and decrease with the increase of laser power. The low concentration of photoexcited electron-hole pairs at small $P_{laser}$ leads to less recombination probability and stronger acousto-electric drag effect. Such behavior is useful in weak-light detection. As shown in Fig. S12, the responsivity of different calculation methods is compared, which shows negligible difference. From the perspective of conversion efficiency, we calculate EQE (Fig. 4c) from the experiment and the simulation with the change of acoustic intensity. The EQE of all the devices grows appreciably as acoustic intensity rises while the growth rate slows down. As a result, the EQE of the devices with the best photoresponse can be promoted to a high value over 60%. However, we note that the operation procedure of ADPVE requires the application of SAW, which involves additional energy input other than the photon alone. According to calculation and experiments (Section 4 of Supplementary Information), the energy loss of SAW travelling through a layered sample is estimated to be less than 10% and the possible influence from SAW attenuation at electrodes is excluded. Enlarging the device channel and recycling the output RF power could help to enhance the device efficiency in terms of the energy consumption tradeoff between the input SAW and the obtained photocurrent.

Fig. 4d shows the detected signals at a function of time when a pulse of SAW of 300 μs is applied. The top and bottom panels of Fig. 4d represent the input RF power of SAW and the open-circuit photovoltage response, respectively. We fabricated a photodetector array with three two-terminal $MoSe_2$ samples on the same SAW device (see Fig. S13). The $MoSe_2$ samples were in the longitudinal configuration with different distance away from the IDT. The difference in distance among device positions then leads to different delay time of acousto-drag photovoltaic signals with respect to the excitation time of SAW pulse. Fig. S13 summarizes the time delay $\Delta t$ of different $MoSe_2$ photodetectors versus the distance from the launching IDT. It well fits to the theoretical relation (the red dashed line) given by the sound velocity, which verifies the transport of photoexcited pairs under SAW. It offers a new way of spatial-resolved detection by tracking the time-domain response.

To compare the device performance with other systems with nonlocal directional photocurrent, we summarize the photocurrent density of different materials in Fig. 4e. Due to the linear dependence of the photocurrent on light power density, we mark the working region with shallow bars. The photocurrent density given by the $MoSe_2/LiNbO_3$ device (SAW-ON state) in this study stands out in the scaling plot with laser power density. The comparison demonstrates ADPVE presented in this work can help to achieve efficient optoelectrical generation without additional symmetry restriction on the material. A detailed comparison of device performance among several closely-related representative systems are provided in Table S2.

The finite-element method was further applied to simulate the optoelectronic characteristics of $MoSe_2$ under SAW on $LiNbO_3$ substrate. The geometrical parameters were set similarly to those in the experiment. In order to define the model, it was necessary to apply structural and electrical boundary conditions, such as fixed underside of the $LiNbO_3$ substrate as we assume that the SAW damps within few wavelengths from the surface. Then the modulation of $MoSe_2$ energy band structure by SAW was extracted (Fig. S7). The photocurrent density induced by the above conditions was calculated under the model of Semiconductor-Electromagnetic waves coupling and Ideal Schottky contacts. Afterwards, the total current flowing through the contact was obtained by integration. The input laser power was set at 0.5 μW and the frequency of SAW was 300 MHz. The acoustic power dependence of the quantum efficiency from simulation generally agrees with the obtained experimental data as shown in Fig. 4c.

The giant nonlocal photocurrent generation by SAW may potentially lead to broad application prospect in the field of photodetection and energy harvesting. The current work mainly represents a proof-of-concept demonstration of ADPVE by piezoelectric integration, since the pristine $MoSe_2$ device is not a well-designed photodetector or solar cell due to the small and compensating Schottky barriers. Further improvement can be made by adopting more efficient photocurrent extraction approaches such as *p-n* junctions, Schottky junctions with larger barrier height, as well as other BPVE mechanisms. Since SAW is a well-developed technology, various configurations of IDTs such as focused IDTs, nonreflective IDTs may also be applied to increase the acoustic intensity and reduce the energy consumption of SAW excitation. More importantly, the two key components of photo-electric conversion: the electron-hole pair generation part (typically semiconductors with gap size closed to the targeted wavelength) and the photocurrent extraction part (typically junctions with large build-in field) are now spatially separated and independent of each other. It allows for further optimization on each part without their interior tradeoff, which may be useful for designing new infrared detectors. Owing to the effective suppression of carrier recombination, the integration of SAW can be a potential solution for improving efficiency in narrow-gap-semiconductor or semimetal-based infrared photodetectors.

In conclusion, we report the realization of ADPVE by integrating two-dimensional semiconductor $MoSe_2$ on piezoelectric substrates $LiNbO_3$. Similar to BPVE, it manifests as the emergence of zero-bias photocurrent in the bulk of a homogenous material. Instead of relying on inversion symmetry breaking, ADPVE results from the charge dragging effect of SAW, which generates nonlocal photocurrent in the bulk far away from Schottky barrier. Meanwhile, SAW also leads to the spatial separation of photoexcited electron-hole pairs, further contributing to photocurrent enhancement. We demonstrate ADPVE by piezoelectric integration as a new design for generating zero-bias nonlocal photocurrent, which may be used to help design sensitive photodetectors and high-efficient solar cells.

**Acknowledgments**
C.Z. was sponsored by the National Key R&D Program of China (Grant No. 2022YFA1405700), the National Natural Science Foundation of China (Grant No. 92365104 and 12174069), and Shuguang Program from the Shanghai Education Development Foundation. W.S. was supported by the National Natural Science Foundation of China (Grant No.12274090) and the Natural Science Foundation of Shanghai (Grant No.22ZR1406300). Part of the sample fabrication was performed at Fudan Nano-fabrication Laboratory. We thank Prof. W. Yu, Prof. P. Wang and Prof. Y. Liu for helpful discussion.

**Author contributions**

C.Z. conceived the ideas and supervised the overall research. J.G. and Y.M. fabricated and characterized the device with the help of C.Z., Y.W., H.G., W.S., and X.J. J.G. and J.M. carried out the scanning photocurrent measurements. H.C. and J.W. performed the simulation. C.Z. and J.G. analyzed the data. J.G. and C.Z. wrote the paper with help from all other co-authors.

**Competing financial interests**

A related patent application has been filed by Fudan University with C.Z., J.G., Y.M., J.M. and H.C. as inventors.

**Data and Code availability**

Source data are provided with this paper. All other data that support the findings of this study are available from the corresponding authors upon reasonable request. The codes used for simulation are available from the corresponding authors upon reasonable request.

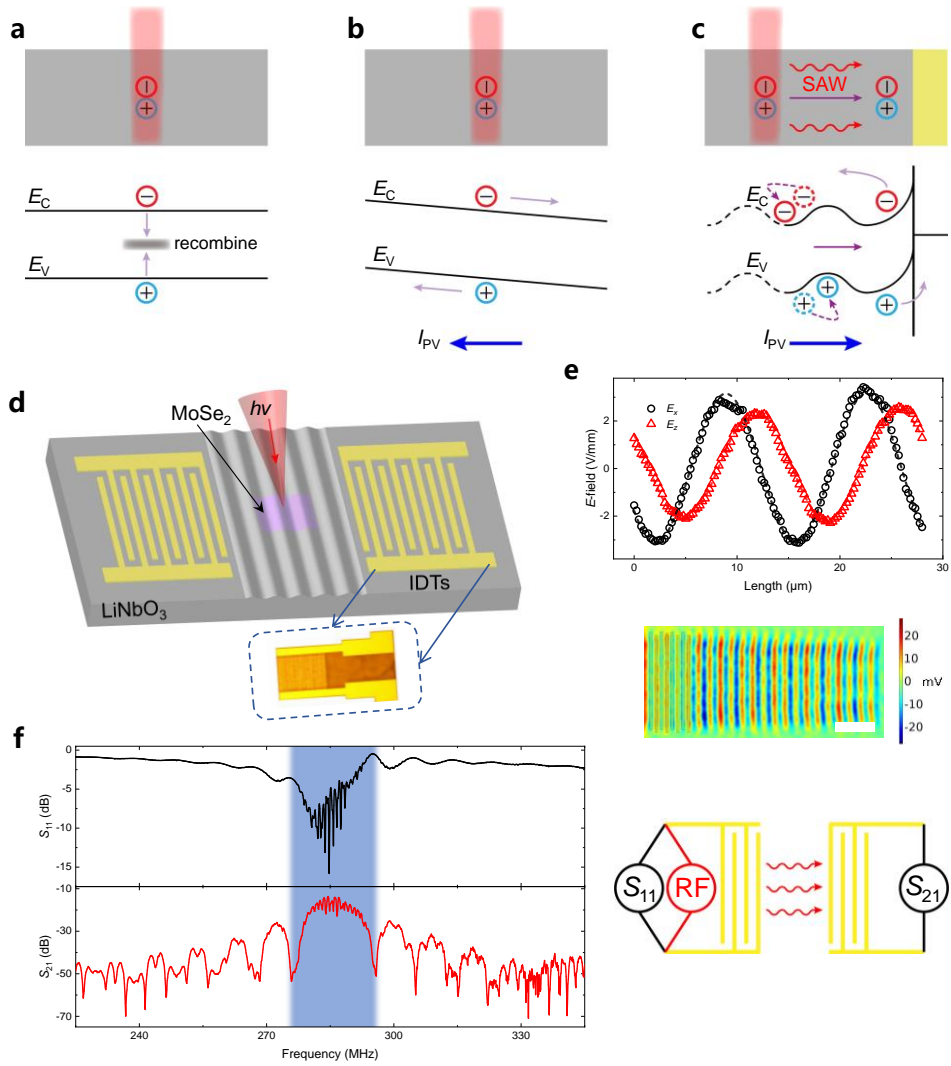

**Fig. 1: Illustrations of different photocurrent generation mechanisms and MoSe$_2$/LiNbO$_3$ device configuration. a**, In unbiased semiconductors, the photoexcited electron-hole pairs recombine through recombination centers without generating directional photocurrent. **b**, In biased semiconductors, the photoexcited pairs are separated by electric fields and form photocurrent by band bending. **c**, In semiconductors with periodical band modulation by SAW, the photoexcited pairs are spatially separated to the minima of conduction band and the maxima of valence band, and collectively dragged along the travelling direction of SAW. The spatial separation effectively prevents them from recombination. Photocurrent is extracted through a local Schottky barrier. **d**, Schematic illustration of MoSe$_2$ placed on a LiNbO$_3$ substrate. The IDT pairs generates SAW travelling through the MoSe$_2$ flake. The inset is the optical image of one IDT. The yellow color is the gold; the brown color is the LiNbO$_3$ substrate. **e**, The in-plane and out-of-plane piezoelectric fields ($E_x$, $E_z$) distribution by simulation and the piezoelectric potential profile of the SAW generated by the IDT. The acoustic $E$-potential travels forward uniformly at a certain period, which matches with the sound speed and frequency. Scale bar, 50 μm. The input RF power is 1 mW. **f**, $S$-parameters ($S_{11}$, $S_{21}$) of a typical device with a center frequency around 280 MHz and schematic illustration of the measurement. The RF source generates SAW while two RF meters measure the reflection power and transmission power received by the other IDT respectively.

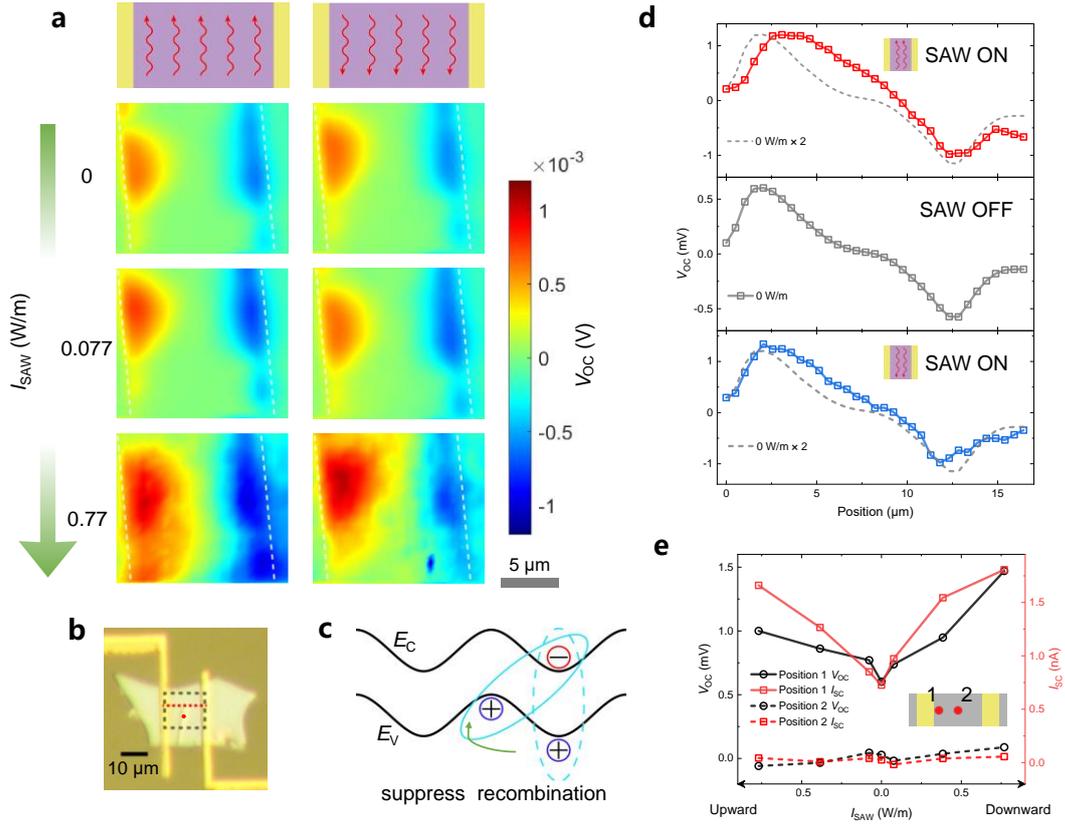

**Fig. 2: Photoexcited electron-hole pair separation effect induced by SAW. a**, Open-circuit photovoltage ($V_{OC}$) mapping series under a laser power of 0.25 μW. The red wavy arrows indicate the directions of SAW propagation, which are vertical to the direction of the photovoltage gradient. The white lines outline the Au electrodes boundary. The gradual green arrow represents the acoustic intensity of SAW, which increases from 0 to 0.77 W/m. Photovoltages increase with SAW at the region of the Schottky junction, the signs of which remain unchanged when the SAW direction is reversed. Scale bar, 5 μm. **b**, Optical microscope image of Device 1, with the black dotted line indicating the scanning region and the red dotted line indicating the scanning line. The red circle represents the exact laser size. Scale bar, 10 μm. **c**, Illustration of the spatial separation effect of photoexcited electrons and holes when applying SAW. **d**, The line-profile analysis with a laser power of 0.25 μW, corresponding to the red dotted line marked in Fig. 2b. The acoustic intensity of SAW is 1.54 W/m for both top and bottom panels. The SAW propagation direction represented by inset in top panel is upward, and which in bottom panel is downward. $V_{OC}$ without SAW is magnified at 2 times in both top and bottom panels. **e**, The open-circuit photovoltage ($V_{OC}$, black) and short-circuit photocurrent ($I_{SC}$, red) variations with the acoustic intensity and SAW propagation direction. The inset represents the position of the laser with a power of 0.25μW. We choose the $V_{OC}$ and $I_{SC}$ of position 1 of the MoSe$_2$ flake and position 2 in Fig. 2b (the region of the Schottky junction) for comparison. The arrows of acoustic intensity refer to the opposite propagation direction of SAW.

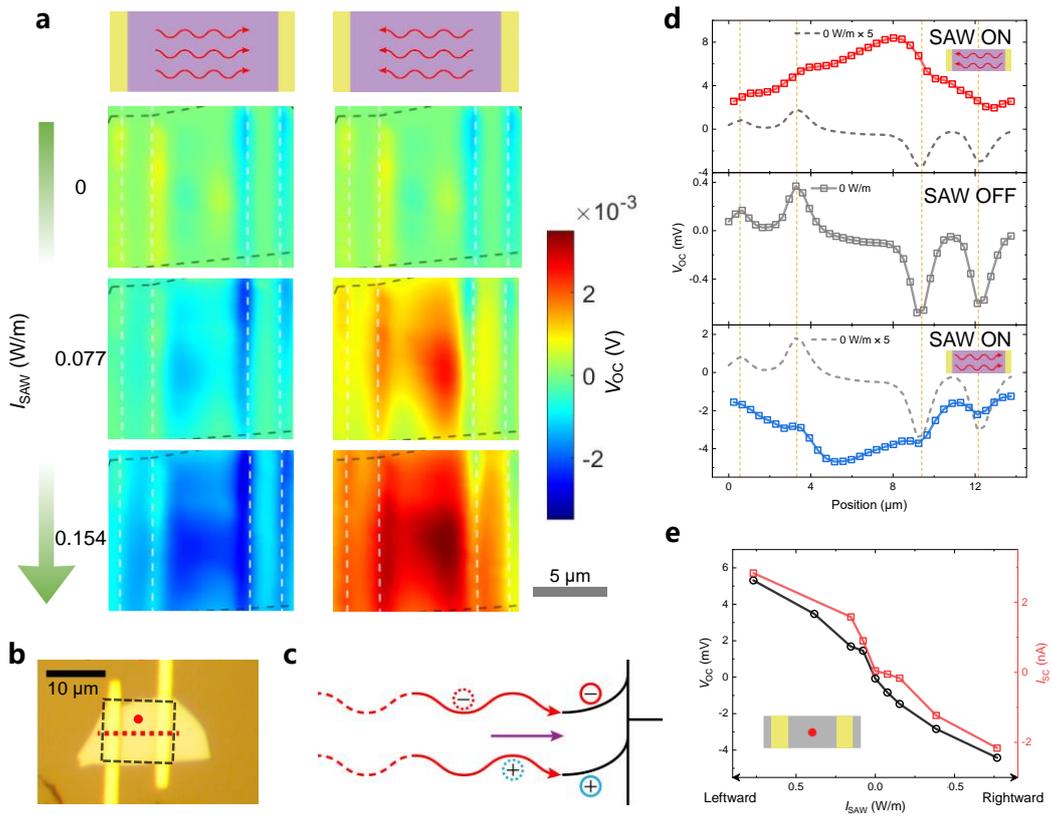

**Fig. 3: Acousto-electric drag effect of photoexcited electron-hole pairs induced by SAW. a**, Open-circuit photovoltage ($V_{OC}$) mapping series under a laser power of 0.39 μW. The red wavy arrows indicate the directions of SAW propagation, which are parallel to the direction of the photovoltage gradient. The gradual green arrow represents the acoustic intensity of SAW, which increases from 0 to 0.154 W/m. The black dashed lines outline the MoSe$_2$ flake and the white dashed lines outline the Au electrodes. Photovoltages increase with acoustic intensity as the laser is focused in the center region between the contact electrodes, and switches signs when the SAW direction is reversed. Scale bar, 5 μm. **b**, Optical microscope image of Device 2, with the black dotted line indicating the scanning region and the red dotted line indicating the scanning line. The red circle represents the exact laser size. Scale bar, 10 μm. **c**, Illustration of the acousto-electric drag effect of carriers when applying SAW. **d**, The open-circuit photovoltage ($V_{OC}$, black) and short-circuit photocurrent ($I_{SC}$, red) variations with acoustic intensity and SAW propagation direction. The arrows of acoustic intensity refer to the opposite propagation direction of SAW. The inset represents the position of the laser with a power of 0.40 μW. **e**, The open-circuit photovoltage ($V_{OC}$, black) and short-circuit photocurrent ($I_{SC}$, red) variations with the acoustic intensity and SAW propagation direction. The inset represents the position of the laser with a power of 0.40 μW. The arrows of acoustic intensity refer to the opposite propagation direction of SAW.

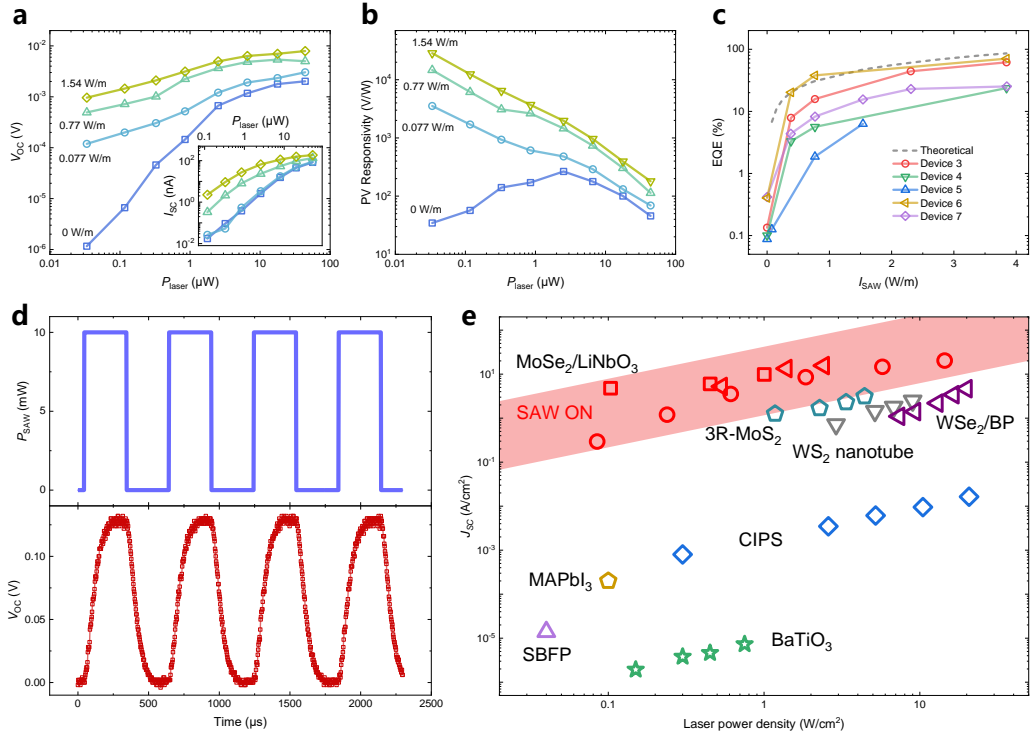

**Fig. 4: Optoelectronic characterizations of ADPVE in the MoSe$_2$/LiNbO$_3$ device. a**, $V_{OC}$ variation with laser power ($P_{laser}$) at different acoustic intensity. The inset is $I_{SC}$ variation with laser power at different acoustic intensity. **b**, The PV responsivity verse laser power ($P_{laser}$) is presented. The acoustic intensity ranges from 0 W/m to 1.54 W/m in both situations. The corresponding laser power intensity is from 0.024 W/cm$^2$ to 93 W/cm$^2$. **a** and **b** are measured from Device 5. **c**, External quantum efficiency variation with acoustic intensity. The corresponding laser power ranges from 0.10 μW to 0.65 μW. The gray dashed line is the simulated result. **d**, Open-circuit voltage response for Device 11-1. The upper panel is the input RF power of SAW and the bottom panel is the open-circuit voltage. The pulse duration is 300 μs and the period is 600 μs. **e**, Experimental photocurrent density for various systems with nonlocal directional photocurrent. The presented data includes MoSe$_2$/LiNbO$_3$ in this study (Device 2-red circle, Device 3-red square and Device 4-red triangle), WS$_2$ nanotube[6], CIPS[20], BaTiO$_3$[47], SBFP[48], 3R-MoS$_2$[5], WSe$_2$/BP[9], and MAPbI$_3$[49].